\pdfoutput=1

\documentclass[aps,prl,twocolumn,showpacs,superscriptaddress,a4paper,
    floatfix]{revtex4}
\usepackage[latin1]{inputenc}
\usepackage{graphicx}
\usepackage{amsmath,amssymb}
\usepackage{hyperref}
\usepackage{datetime}

\begin{document}

\title{Local observation of antibunching in a trapped Fermi gas}

\author{Torben Müller}
\affiliation{Institute for Quantum Electronics, ETH Zurich, 8093
Zurich, Switzerland}
\author{Bruno Zimmermann}
\affiliation{Institute for Quantum Electronics,  ETH Zurich, 8093
Zurich, Switzerland}
\author{Jakob Meineke}
\affiliation{Institute for Quantum Electronics,  ETH Zurich, 8093
Zurich, Switzerland}
\author{Jean-Philippe Brantut}
\affiliation{Institute for Quantum Electronics,  ETH Zurich, 8093
Zurich, Switzerland}
\author{Tilman Esslinger}
\email{esslinger@phys.ethz.ch} \affiliation{Institute for Quantum
Electronics, ETH Zurich, 8093 Zurich, Switzerland}
\author{Henning Moritz}
\affiliation{Institute for Quantum Electronics,  ETH Zurich, 8093
Zurich, Switzerland} \affiliation{Institut für Laser-Physik,
Universität Hamburg, 22761 Hamburg, Germany}

\date{\pdfdate}

\begin{abstract}
Local density fluctuations and density profiles of a Fermi gas are
measured in-situ and analyzed. In the quantum degenerate regime, the
weakly interacting $^6$Li gas shows a suppression of the density
fluctuations compared to the non-degenerate case, where atomic shot
noise is observed. This manifestation of antibunching is a direct
result of the Pauli principle and constitutes a local probe of
quantum degeneracy. We analyze our data using the predictions of the
fluctuation-dissipation theorem and the local density approximation,
demonstrating a fluctuation-based temperature measurement.
\end{abstract}


\maketitle


A finite-size system in thermodynamic equilibrium with its
surrounding shows characteristic fluctuations, which carry important
information about the correlation properties of the system. In a
classical gas, fluctuations of the number of atoms contained in a
small sub-volume yield a Poisson distribution, reflecting the
uncorrelated nature of the gas. An intriguing situation arises when
the thermal de Broglie wavelength approaches the interparticle
separation and the specific quantum statistics of the constituent
particles becomes detectable. For bosons, positive density
correlations build up, until Bose-Einstein condensation occurs, as
measured in Hanbury Brown-Twiss (HBT) experiments
\cite{hanbury_brown_test_1956,
Baym_1998,Yasuda_1996,altman_probing_2004,
Oettl_2005,folling_spatial_2005,schellekens_hanbury_2005}. The
effect of bunching also manifests itself in enhanced density
fluctuations in real space \cite{esteve_observations_2006}. In
contrast, fermions obey the Pauli principle. This gives rise to
anti-correlations, which have been observed in HBT experiments
\cite{henny_fermionic_1999,oliver_hanbury_1999,iannuzzi_direct_2006,rom_free_2006,
jeltes_comparison_2007}, and are expected to squeeze density
fluctuations below the classical shot noise limit \cite{belzig_2007}.
Moreover, for trapped fermions, the reduction of fluctuations varies
in space, reaching a maximum in the dense center of the cloud, which
should be accessible to a local measurement.

In this letter we report on high-resolution in-situ measurements of
density fluctuations in an ultracold Fermi gas of weakly interacting
$^6$Li atoms. We extract the mean and the variance of the density
profile from a number of absorption images recorded under the same
experimental conditions. Our measurements show that the density
fluctuations in the center of the trap are suppressed for a quantum
degenerate gas as compared to a non-degenerate gas. We analyze our
data using the fluctuation-dissipation theorem, which relates the
density fluctuations of the gas to its isothermal compressibility.
This allows us to extract the temperature of the system
\cite{zhou_universal_2009, esteve_observations_2006,
gemelke_in_2009}.

\begin{figure}[b!]
    \includegraphics[width=0.47\textwidth]{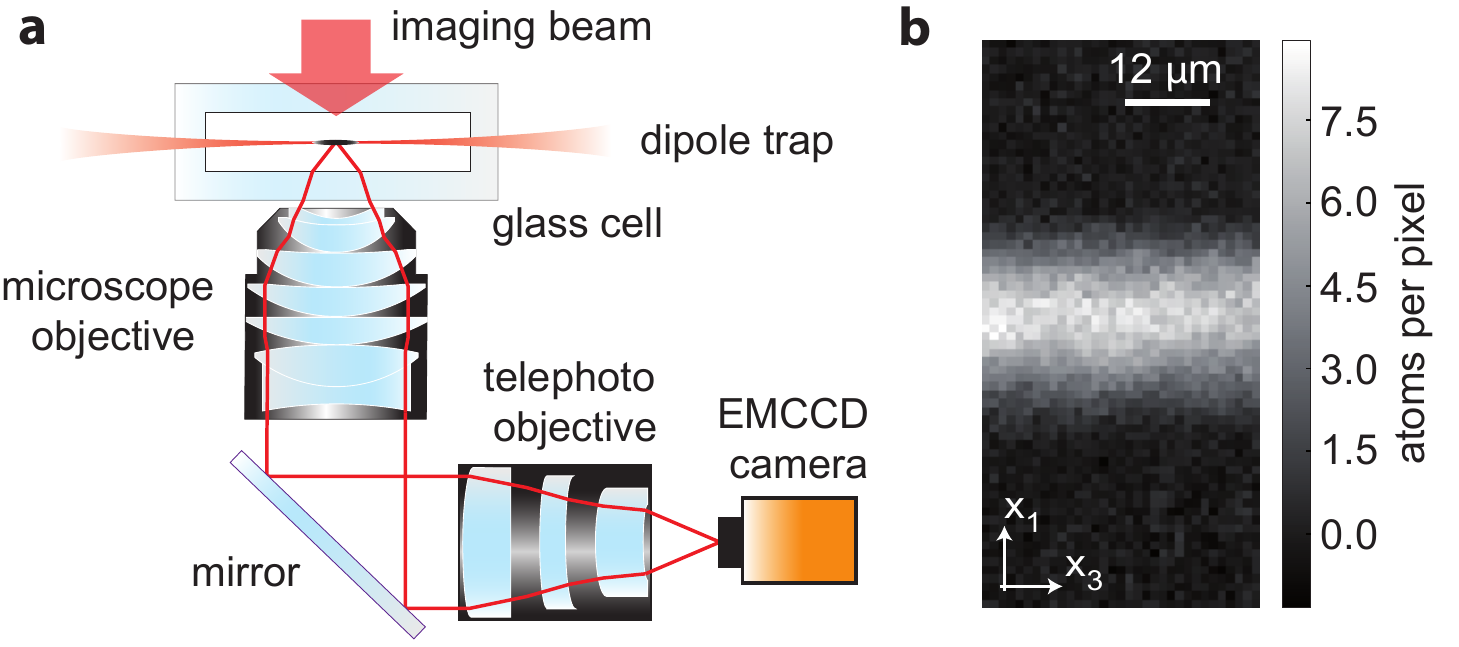}
    \caption{(a) Setup for high-resolution imaging of the trapped $^6$Li 
gas. The
    shadow cast by the atoms held in the dipole trap is imaged through
    the microscope objective and a telephoto objective onto an EMCCD
    chip. The resolution ($1/e^2$-radius) is $1.8\,\mathrm{\mu m}$ 
\cite{depth} at a
    wavelength of $671\,\mathrm{nm}$ and the magnification is $54$. (b) 
Density
    distribution (atoms per pixel) of the trapped atoms obtained by 
averaging over $20$
    realizations. The effective pixel size measures $1.2\,\mathrm{\mu m}$
    \cite{bin}, the maximum optical density is $\sim 2$. The images which we 
evaluate are
    $\sim 40 \,\mathrm{\mu m}$ wide showing the center of the cloud, which 
has a total length of $500\,\rm{\mu m}$.}
    \label{fig:microsetup}
\end{figure}

We first describe the experimental procedure to obtain a quantum
degenerate gas of about $6 \times 10^4$ $^{6}$Li atoms equally
populating the two lowest hyperfine states. Following the method described 
in \cite{jochim_bose-einstein_2003},
the atoms are loaded into an optical dipole trap created by a far
off-resonant laser with a wavelength of $1064\,\mathrm{nm}$, focused
to a $1/e^2$-radius of $(22\pm1)\,\mathrm{\mu m}$ \cite{errors}. The
cloud is then optically moved \cite{gustavson_transport_2001} into a
glass cell that provides high optical access, see Fig.
\ref{fig:microsetup}(a). In the glass cell, forced evaporation is
performed by reducing the trap power from initially $2\,\mathrm{W}$
to $4.7\,\mathrm{mW}$. During evaporation a homogeneous magnetic
field of $300 \,\mathrm{G}$ is applied to set the s-wave scattering
length $a$ for inter-state collisions to $-300\,a_{0}$, where
$a_{0}$ is the Bohr radius. The magnetic field is then ramped to
$475 \,\mathrm{G}$ in $150\,\mathrm{ms}$, changing $a$ to
$-100\,a_{0}$, and finally the power of the trapping beam is
increased to $10 \,{\rm mW}$ in $100\,\mathrm{ms}$. Alternatively,
we prepare the lithium gas at temperatures above quantum degeneracy.
For this, we evaporate to $50 \,{\rm mW}$ before recompressing to
$100 \,{\rm mW}$, followed by a $100 \,{\rm ms}$ period of
parametric heating. In both cases, the cloud is allowed to
thermalize for $350\,\mathrm{ms}$ before an absorption image is
taken. The gas is weakly interacting since $|k_F a|< 10^{-2}$, with
$k_F$ the Fermi wavevector.

Our imaging setup is sketched in Fig. \ref{fig:microsetup}(a). The
probe light, resonant with the lowest hyperfine state of the
$|2S_{1/2}\rangle$ to $|2P_{3/2}\rangle$ transition, is collected by
a high-resolution microscope objective and imaged on an
electron-multiplying CCD (EMCCD) chip. The atoms are illuminated for
$8\,{\rm \mu s}$, each atom scattering about 20 photons on average.
Fig. \ref{fig:microsetup}(b) shows the average density distribution
obtained in $20$ experiments.

We now present our procedure for extracting the spatially resolved
variance of the atomic density. The position of each pixel in the imaging
plane of the camera defines a line of sight intersecting with the atomic
cloud. Correspondingly, each pixel \cite{bin}, having an effective area
$A$, determines an observation volume in the atomic cloud along this line
of sight. At low saturation, the transmission $t$ of the probe light
through an observation volume containing $N$ atoms reads $t =
e^{-\sigma\cdot N/A}$, where $\sigma$ is the photon absorption cross
section. As a consequence, for small Gaussian fluctuations of the atom
number, the {\it relative} fluctuations of the transmission coefficient
are equal to the {\it absolute} fluctuations of the optical density and
are thus directly proportional to the number fluctuations:
\begin{equation}\label{eq:deltan2}
    \frac{\delta t^2}{\langle t\rangle^2} = \frac{\sigma^2}{A^2}\delta
N^2, \end{equation} where $\delta t^2$, $\langle t\rangle$ and
$\delta N^2$ are the variance and the mean of the transmission
coefficient, and the variance of atom number, respectively.

Experimentally, repeated measurements of identically prepared clouds
provide us with a set of count numbers $C$ for each pixel, i.e. each
observation volume, corresponding to a certain number of incoming
photons registered by the EMCCD. Typically we register $\sim 1300$
counts, corresponding to $\sim 130$ photons at the position of the atoms. We 
then compute the
variance $\delta C ^2$ and mean $\langle C \rangle$. The relative
noise of the counts and the relative noise of the transmission are
related by:
\begin{equation}
    \label{eq:deltac2overc2} \frac{\delta C^2}{\langle C\rangle^2}=
    \frac{2 g }{\langle C\rangle} + \frac{\delta
    t^2}{\langle t\rangle^2}.
\end{equation} Here, $g$ is the gain of the camera for converting
photoelectrons to counts. The first term is the contribution of
photon shot noise while the second term is the contribution of
atomic noise. The factor $2$ in the photon shot noise term is caused
by the electron-multiplying register \cite{basden_photon_2003}. We
extract the contribution of the atoms to the relative fluctuations
of the counts, by subtracting photon shot noise on each pixel
according to (\ref{eq:deltac2overc2}). This requires the value of
$g$ (typically $\sim 15$), which we determine from the linear
relationship between the variance and the mean of the number of
counts in a set of repeated measurements. The atom number
fluctuations are subsequently obtained from (\ref{eq:deltan2}). At
this stage no division by a reference image has been performed,
avoiding this source of noise.

To reduce technical noise adding to these fluctuations, we reject
images showing the largest deviations of total atom number or cloud
position \cite{post}, which amounts to excluding about $30\%$ of the
images. The remaining shot-to-shot fluctuations of the total atom
number $\delta N_{\rm tot}^2$ are taken into account by further
subtracting the quantity $ \delta N_{\rm tot}^2/N_{\rm tot}^2
\langle N \rangle ^2 $, which is less than $<2\%$ of $N$
\cite{esteve_observations_2006}. Total probe intensity variations
from shot to shot are below $0.5\,\%$. Applying this algorithm to
each pixel of the images yields a local measurement of the variance
of the atom number.

The mean atom number per pixel is calculated by dividing the mean
transmission profile by the mean of reference images taken without
atoms after each shot, thus averaging shot-noise before division.
The values for variance and mean, obtained by applying the above
procedures, are then averaged along equipotential lines of the trap.
These lines deviate from horizontal lines ($x_3$-axis) in our images
by less than half a pixel ($0.6\,{\rm \mu m}$).

\begin{figure}[htb]
    \includegraphics[width=0.47\textwidth]{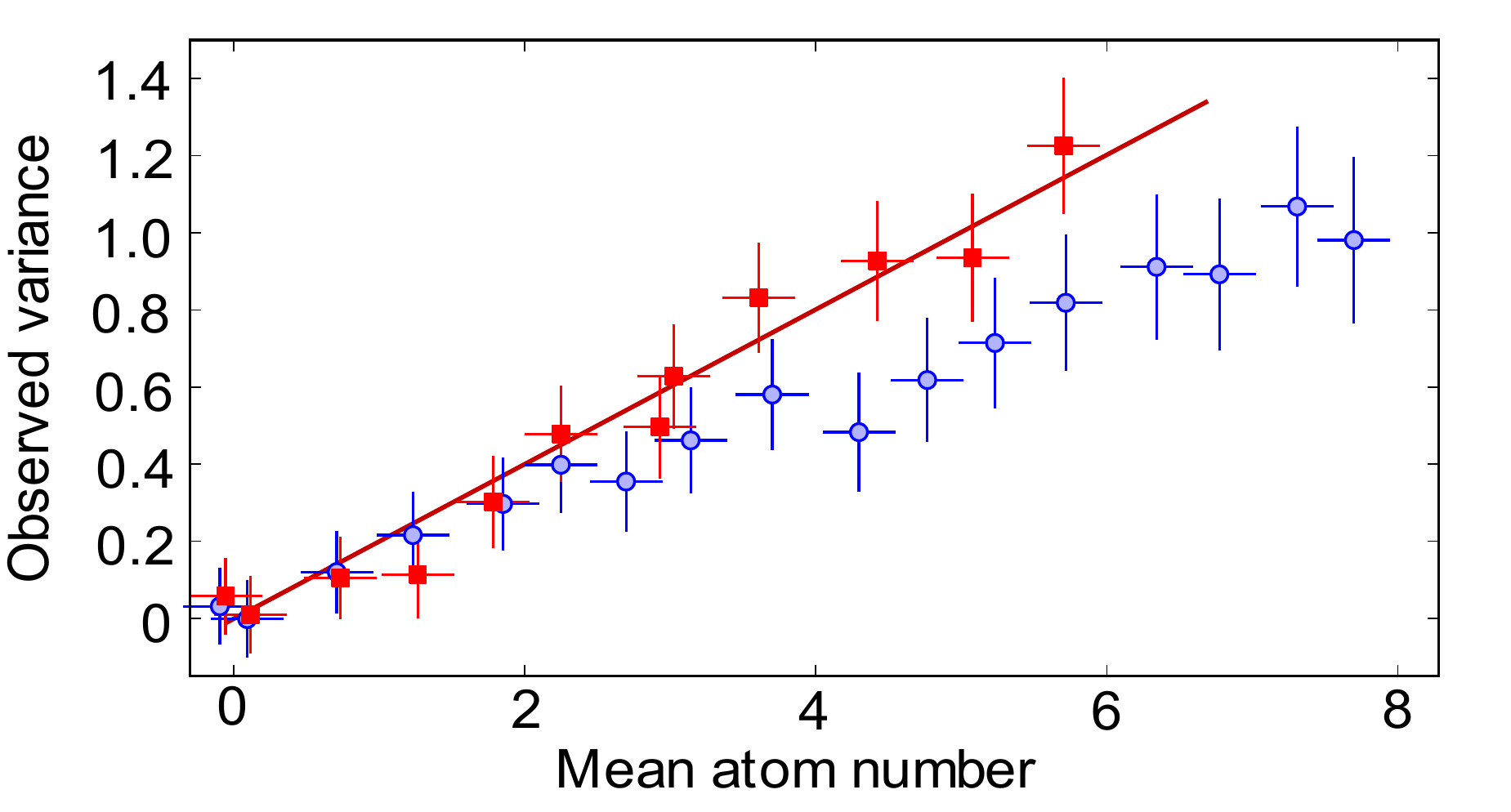}
    \caption{Observed variance versus mean of the atom number detected on a
    pixel. Red squares show the data for a non-degenerate and blue circles
    for a quantum degenerate gas. The solid red line is a linear fit to the 
non-degenerate gas,
    yielding a slope of $0.20\pm0.02$. For the data shown, $80$ experiments 
were performed, $60$ for the degenerate
    case and $20$ for the non-degenerate case. About 30$\%$ of the
    experiments were excluded.
    The error bars shown are estimated from the subtraction of photon shot 
noise which
    is the dominant contribution.}
    \label{fig:varianceVSmean}
\end{figure}

Fig. \ref{fig:varianceVSmean} shows the observed variance of the
atom number plotted against the mean atom number detected on a
pixel. One set of data was taken for a gas at temperatures above
quantum degeneracy (red squares) and another set of data for a
quantum degenerate gas (blue circles). Above quantum degeneracy, the
observed variance is found to be proportional to the mean number of
atoms. The linear behavior confirms that the fluctuations originate
from atomic shot noise.

To quantitatively understand the slope of the noise curve, which is
fitted to be $0.20\pm0.02$, two main effects have to be considered.
These reduce the observed variance and explain the deviation from a
slope of one, which would be expected for the full shot noise.
First, the effective size of the pixel is of the order of the
resolution of the imaging system. As a consequence, the observed
noise is the result of a blurring of the signal over the neighboring
pixels. This effect is also observed in
\cite{esteve_observations_2006,gemelke_in_2009}, and explains a
reduction factor of $0.22$ \cite{binsize}. Second, the probe light
intensity used for the detection is ($15\pm1)\%$ of the saturation
intensity. This leads to a reduction of the photon absorption cross
section due to saturation by $0.95$ and due to the Doppler-shift by
about $0.9$. Together, these effects lead us to expect a slope of
about $0.19$, in good agreement with the observations.

We now turn to the data taken for the quantum degenerate gas (blue
circles in Fig. \ref{fig:varianceVSmean}). At low densities, the
variance is again found to be proportional to the mean density. For
increasingly higher densities, we observe a departure from the
linear behavior and the density fluctuations are reduced compared to
the shot noise limit seen for the non-degenerate gas. This is a
direct consequence of the Pauli principle which determines the
properties of a quantum degenerate Fermi gas. One can think of the
Pauli principle as giving rise to an interatomic "repulsion", which
increases the energy cost for large density fluctuations. This is
similar to the case of bosonic systems with strong interparticle
interactions, where observations have shown a reduction of density
fluctuations \cite{esteve_observations_2006} and squeezing of the
fluctuations below the shot noise limit
\cite{gemelke_in_2009,esteve_squeezing_2008}.

\begin{figure}[htb]
    \includegraphics[width=0.47\textwidth]{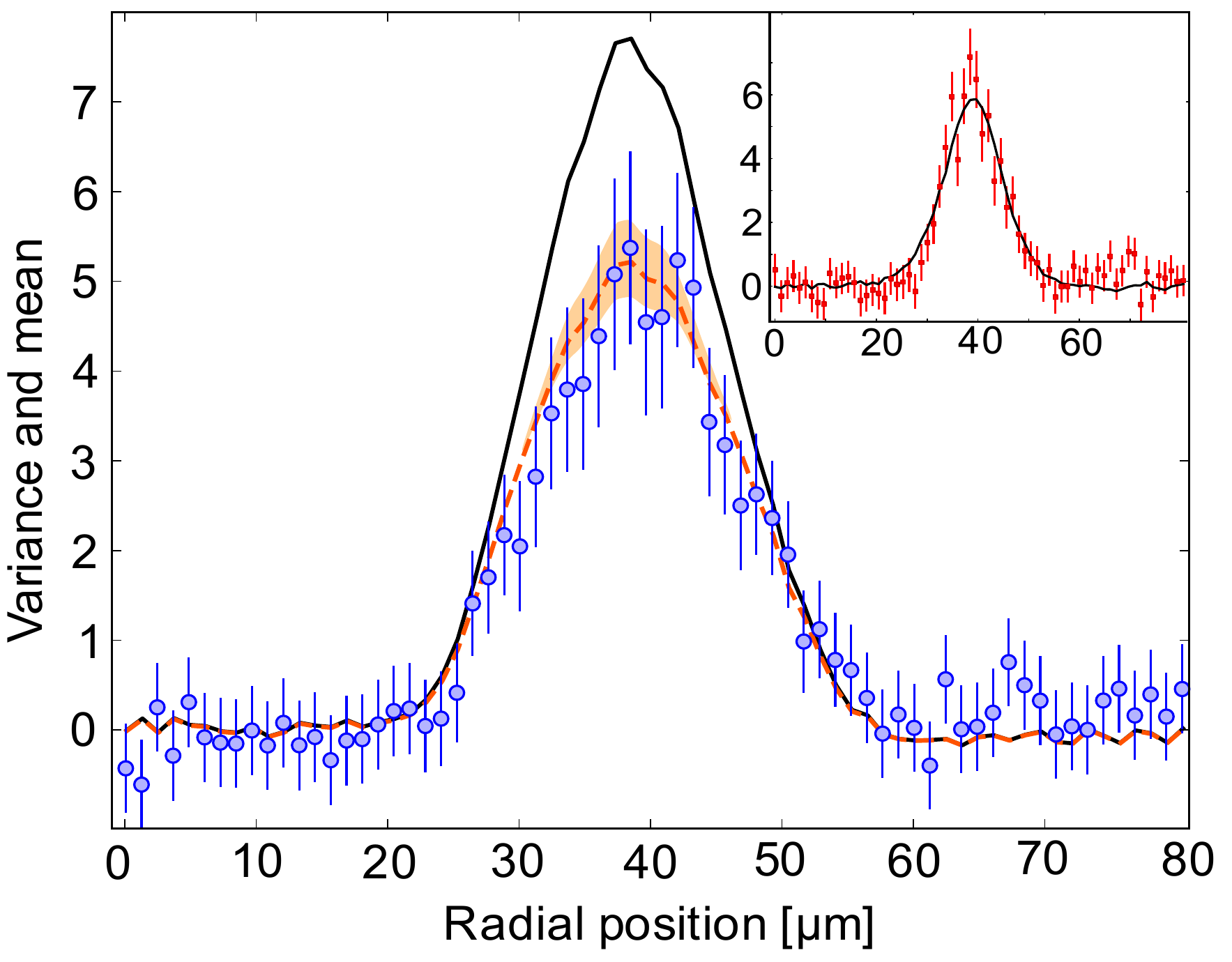}
    \caption{Spatially resolved measurement of antibunching. The
    black line shows the mean atom number and the blue circles the
    corresponding variance along the $x_1$-axis for a degenerate gas.
    The variance is rescaled using the slope fitted in Fig. 
\ref{fig:varianceVSmean}.
    Error bars are estimated from the subtraction of photon shot noise, 
which is the
    dominant contribution. The dashed line shows the variance derived from 
theory.
    The shaded region indicates the uncertainty originating from 
uncertainties in
    the trap parameters. The inset shows corresponding data for a 
non-degenerate gas.}
    \label{fig:results}
\end{figure}

In contrast to previous measurements on antibunching
\cite{rom_free_2006, jeltes_comparison_2007}, we have measured
density fluctuations in a spatially resolved way. For the
construction of Fig. \ref{fig:varianceVSmean}, we have averaged the
observed variance for regions of equal mean density, whereas Fig.
\ref{fig:results} shows the variance (blue circles) and the mean
(black line) of the atom number as a function of the radial position
in the trap for a quantum degenerate gas. While the variance is
proportional to the mean in the wings, at low density, we observe a
reduction of the variance by about $2\,{\rm dB}$ close to the
center, at higher density. The inset shows data for a non-degenerate
gas; in both cases the variance has been rescaled using the slope
fitted in Fig. \ref{fig:varianceVSmean}. The reduction of
fluctuations is a direct indication of the level of quantum
degeneracy of the gas. The larger the average occupation of a single
quantum state, the more the effect of the Pauli principle becomes
important and fluctuations are consequently suppressed. Fig.
\ref{fig:results} thus represents a direct measurement of the local
quantum degeneracy, which is larger in the center of the cloud than
in the wings.

To understand this quantitatively, we describe the atoms contained
in an observation volume in terms of the grand-canonical ensemble
with a local chemical potential fixed by assuming local density
approximation. For a non-interacting gas, the ratio of mean atom
number and its variance is determined by the fugacity $z$ of the
system. This leads to the equation
\begin{equation}\label{eq:rat}
\frac{\delta N^2}{\langle N \rangle} =
\frac{\int\mathrm{Li}_{1/2}(-z(x_1,x_2))dx_2}{\int\mathrm{Li}_{3/2}(-z(x_1,x_2))dx_2},
\end{equation}
where $\mathrm{Li}_i$ is the $i$-th polylogarithmic function, $x_1$
and $x_2$ are radial coordinates of the cloud and line-of-sight
integration is performed along the $x_2$-axis.

The dashed line in Fig. \ref{fig:results} is computed using
(\ref{eq:rat}). For the computation we make use of the Gaussian
shape of the trap, the central fugacity of $13_{-4}^{+18}$, obtained
in an independent time-of-flight experiment (see below), and the
experimental density profile. Our description in terms of the
grand-canonical potential assuming an ideal gas reproduces the
experimental data within the error bars  \cite{grand-canonical}.

We now focus on the interpretation of our results using the
fluctuation-dissipation theorem. At thermal equilibrium, density
fluctuations are universally linked to the thermodynamic properties
of the gas through the fluctuation-dissipation theorem, given by
\begin{equation}\label{eq:fluctdiss}
    k_B T \frac{\partial \langle N \rangle}{ \partial \mu} = \delta
    N^2.
\end{equation}
Here $T$ is the temperature of the gas, $\mu$ the chemical potential
and $k_B$ the Boltzmann constant. Since the local density
approximation allows one to assign a local chemical potential to any
position in the trap, it is possible to determine the
compressibility $\frac{\partial \langle N\rangle}{\partial \mu}$
directly from the mean density profiles \cite{gemelke_in_2009}. From
(\ref{eq:fluctdiss}), the ratio of this quantity to the measured
variance profile of the cloud provides a universal temperature
measurement \cite{zhou_universal_2009}.

\begin{figure}[htb]
    \includegraphics[width=0.47\textwidth]{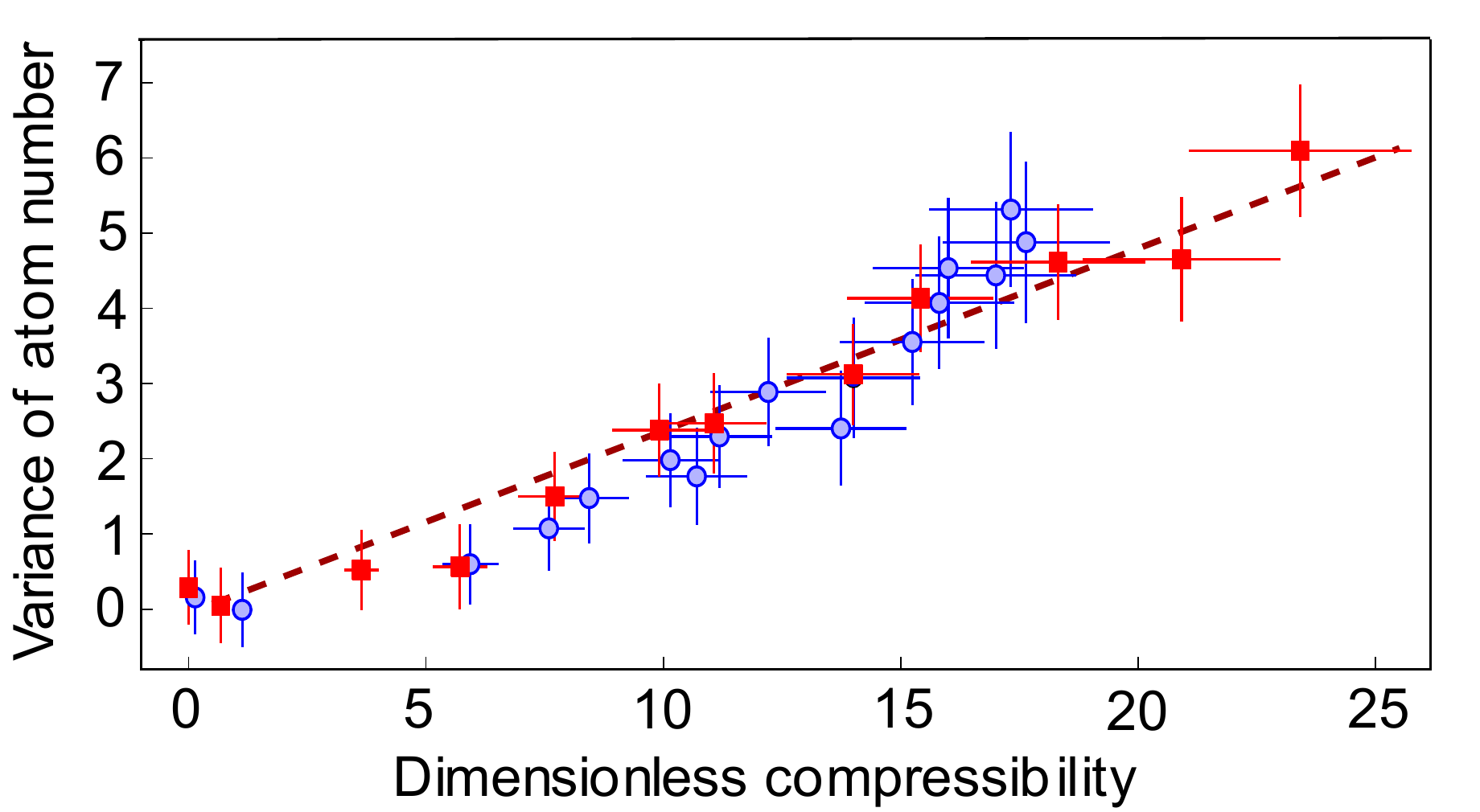}
    \caption{Fluctuation-based temperature measurement. Variance of atom 
number detected on an effective pixel versus
    dimensionless compressibility. The blue circles and red squares show the 
data for the
    quantum degenerate and the non-degenerate case, respectively.
    The dashed red line is fitted to the red squares, giving the temperature 
according to (\ref{eq:fluctdiss}).}
    \label{fig:thermometrydata}
\end{figure}

We apply this procedure to our data by computing the
compressibility, $\frac{\partial \langle N \rangle}{ \partial
\mu}=\frac{\partial \langle N \rangle}{ \partial
x}\left(\frac{\partial \mu }{ \partial x}\right)^{-1}$, where we
take the Gaussian shape of the optical dipole trap into account. To
avoid the problems of numerically differentiating experimental data,
we fit the mean density profile with a linear combination of the
first six even Hermite functions and use the fitted curve as a
measure of the density profile in (\ref{eq:fluctdiss})
\cite{hermite}. Fig. \ref{fig:thermometrydata} shows the variance of
atom number plotted against the dimensionless compressibility $U_0
\frac{\partial \langle N \rangle}{\partial \mu}$, where $U_0$ is the
trap depth. We observe the linear relation described by
(\ref{eq:fluctdiss}) with a slope of $\frac{k_B T}{U_0}$ =
$0.27\pm0.04$ for both data sets, the degenerate and the
non-degenerate. From the physics of evaporative cooling it is
expected that both slopes are the same \cite{luiten_kinetic_1996}.
Using trap depths derived from the measured laser powers, we obtain
temperatures of $(145\pm 31)\,{\rm nK}$ and $(1.10 \pm 0.06)\,{\rm
\mu K}$ for the quantum degenerate and the non-degenerate gas,
respectively.

To assess the quality of this measurement, we have also performed
time-of-flight measurements with clouds prepared under the same
conditions. In this method, we determine the temperatures by fitting the
measured density profiles after free expansion of $1.5 \,{\rm ms}$ ($1
\,{\rm ms}$ for the non-degenerate gas) to the calculated shape of a
non-interacting gas released from a Gaussian trap. This procedure gives us
slightly higher temperatures for the degenerate and the non-degenerate
clouds, which are $(205\pm30)\,{\rm nK}$ ($T/T_F=0.34\pm0.1$) and $(1.6\pm
0.2 )\,{\rm \mu K}$ ($T/T_F=1.9\pm 0.1$), respectively. We attribute the
discrepancy between the two temperature measurements to deviations of the
trapping beam from a Gaussian shape and residual experimental fluctuations
from shot to shot, which both affect the two methods.

In conclusion, we have measured density fluctuations in a trapped
Fermi gas locally, observing antibunching in a degenerate gas. We
have also determined the temperature using the
fluctuation-dissipation theorem \cite{Manz_2010}. In contrast to
bosons, fermions cannot exhibit first-order long range coherence due
to the Pauli principle. However, when a Fermi system enters a
quantum correlated phase, e.g. a superfluid phase, long range
even-order correlations build up \cite{yangODLRO}. The spatially
resolved measurement of density fluctuations, probing second-order
correlations, is thus a natural tool to study strongly correlated
Fermi gases.

We acknowledge discussion with T.L. Ho, M. Greiner and D. Stadler
and financial support from ERC SQMS and FP7 FET-open NameQuam.

After submission of this article, similar results were reported in
\cite{sanner_suppression_2010}.


\begin{thebibliography}{10}
\bibitem{errors}

\bibinfo{note}{Unless otherwise stated, errors are a combination of 
statistical
  errors and uncertainties in the determination of experimental 
parameters.}
\bibitem{bin}

\bibinfo{note}{4x4 software binning of camera pixels define the pixels 
referred
  to in the text. Camera pixels have an effective size (when imaged to the
  object plane) of 300\, nm.}
\bibitem{post}

\bibinfo{note}{Images deviating more than 1.5 (1.0) standard deviations, 
corresponding to $0.72\,\rm{\mu m}$ ($5\,\%$) in the mean
  position (total atom number) are excluded.}
\bibitem{binsize}

\bibinfo{note}{A simulation accounting for the optical resolution was 
performed.
16x16 binning (instead of 4x4) increases the slope to 0.8 agreeing with the 
simulation.}
\bibitem{grand-canonical}

\bibinfo{note}{The Fermi wavelength is of the order of the pixel size. 
However,
  column integration is expected to average out possible correlations.}
\bibitem{hermite}

\bibinfo{note}{The difference of this method with direct numerical 
differentiation
  of data (where possible) is within the error bars.}
\bibitem{depth}

\bibinfo{note}{The resolution of the NA=0.55 microscope objective has been
  artificially reduced with a diaphragm in order to increase the depth of 
field
  to the order of the cloud size.}
\end{thebibliography}

\end{document}